\documentstyle[twoside,fleqn,npb,epsfig,axodraw]{article}
\def\e{{\rm e}}
\newcommand{\be}{\begin{equation}}
\newcommand{\ee}{\end{equation}}
\newcommand{\bea}{\begin{eqnarray}}
\newcommand{\eea}{\end{eqnarray}}

\newcommand{\gm}{\gamma}
\newcommand{\Gm}{\Gamma}

\newcommand{\ep}{\epsilon}

\newcommand{\dd}{\mbox{d}}

\newcommand{\lra}{\leftrightarrow}

\newcommand{\nn}{\nonumber}

\newcommand{\Li}[2]{{\mbox{Li}}_{#1}\left(#2\right)}

\newcommand{\AmS}{{\protect\the\textfont2
  A\kern-.1667em\lower.5ex\hbox{M}\kern-.125emS}}

\title{
\rightline{\normalsize hep-ph/0209295}
\rightline{\normalsize September 2002}
Evaluating double and triple (?) boxes}

\author{ V.A. Smirnov\address{
Nuclear Physics Institute of Moscow State University,\\
Moscow 119992, Russia}%
        \thanks{
Talk  presented at the International Symposium Radcor 2002 and Loops and Legs
2002 (September 8--13, Kloster Banz, Germany).
Supported by  INTAS through grant 00--00313,
the Russian Foundation for Basic Research through project 01--02--16171,
and the Volkswagen Foundation, contract
No.~I/77788.}
}

\begin{document}

\begin{abstract}
A brief review of recent results on analytical evaluation of double-box
{}Feynman integrals is presented. First steps towards evaluation of massless
on-shell triple-box Feynman integrals within dimensional regularization are
described. The leading power asymptotic behaviour of the dimensionally
regularized massless on-shell master planar triple-box diagram in the Regge
limit $t/s \to 0$ is evaluated. The evaluation of the unexpanded master planar
triple box is outlined and explicit results for coefficients
at $1/\ep^j$, j=2,\ldots,6, are presented.
\end{abstract}


\maketitle


\section{Double boxes}

 {}Feynman diagrams with four external lines contribute to many important
physical quantities. They are rather complicated mathematical objects
because they depend on many variables:
internal masses,  Mandelstam variables and
squares of external momenta. The most complicated two-loop diagrams are
the planar double box and non-planar (crossed) double box.
Almost all available analytical results correspond to the massless diagrams.
Ironically, the first result for the massless
double boxes was obtained in the most complicated case, where
all the fours external legs are off-shell, i.e.
$p_i^2\neq 0$ for all $i=1,2,3,4$. This is an elegant analytical result
for the scalar planar master double box (and, moreover, for a general planar
ladder diagram of this type), i.e. for all powers
of propagators equal to one, obtained in \cite{UD}.
However, no other results for the pure off-shell double
boxes (e.g. with dots on some lines and/or with some lines,
other than rungs, contracted) have been derived up to now so that
this result stays unique in the pure off-shell category.

{}For massless double-box diagrams with at least one leg on the mass shell,
i.e. $p_i^2=0$, infrared and collinear divergences appear,
so that one introduces a regularization which is usually chosen to be
dimensional \cite{dimreg}, with the space-time dimension $d$ as a
regularization parameter.
One hardly believes that a regularized double-box diagram
can be analytically evaluated for the general value of the regularization
parameter $\ep=(4-d)/2$, and the evaluation is usually performed
in a Laurent expansion in $\ep$, typically, up to a finite part.

The problem of the evaluation of Feynman integrals
associated with a given graph according to some Feynman rules
is usually decomposed into two parts: reduction of general Feynman integrals
of this class to so-called master integrals (which cannot be simplified
further) and the evaluation of these master integrals.
A standard tool to solve the first part of this problem is
the method of integration by parts (IBP) \cite{IBP} when
one writes down identities obtained by putting to zero various
integrals of derivatives of the general integrand connected
with the given graph and tries to solve a resulting system of equations
to obtain recurrence relations that express Feynman integrals with
general integer powers of the propagators through the master integrals.

The most complicated basic master planar and non-planar on-shell massless
double-box diagram were calculated in \cite{K1,Tausk} by a method based on
Feynman parameters and Mellin--Barnes (MB) representation (see \cite{S4}
for details of the method).
It turns out that it is natural to consider non-planar double boxes as
functions of the three Mandelstam variables $s,t$ and $u$ not
necessarily restricted by the physical condition $s+t+u=0$ which does not
simplify the result.

Reduction procedures for the evaluation of general double-box diagrams,
with arbitrary numerators and integer powers of the propagators
were developed in \cite{SV} in the planar case and in \cite{AGORT} in the
non-planar case.
In \cite{SV}, the first of the two most complicated master integrals
involved is with all powers of propagators equal to one.
As a second complicated master
integral, the authors of \cite{SV} have chosen the diagram with a dot
on the central line.
As was pointed out later \cite{GT}, in practical calculations one runs into
a linear combination of these two master integrals with the coefficient
$1/\ep$, so that a problem has arisen because the calculation of
the master integrals in one more order in $\ep$ looked rather nasty.
Two solutions of this problem immediately appeared. In \cite{GR0},
the authors calculated this very combination
of the master integrals, while in \cite{ATT} another choice of
the master integrals was made:  as a second complicated master
integral, the authors have taken the integral which is obtained
from the first master integral by inserting a specific numerator.
This was a more successful choice because,
according to the calculational experience, no negative
powers of $\ep$ occur as coefficients at these two master integrals.

These  analytical algorithms were  successfully
applied to the evaluation of two-loop virtual
corrections to various scattering processes \cite{appl} in the
zero-mass approximation.

In the case, where one of the external
legs is on-shell, $p_1^2\neq 0$, $p_i^2=0,\;i=2,3,4$,
the planar double box and one of two possible non-planar double-box
diagrams with all powers of propagators equal to one
were analytically calculated in \cite{S2}, as functions of
the Mandelstam variables $s$ and $t$ and the non-zero external momentum
squared $p_1^2$. Explicit
results were expressed through  (generalized) polylogarithms, up to the
fourth order,  dependent on rational combinations of $p_1^2,s$
and~$t$, and a one- and (in the non-planar case) two-dimensional integrals
with simple integrands. To do this, the method based on MB integrals
mentioned above was applied.
These and other master planar and non-planar double boxes with one leg
off-shell were evaluated in \cite{GR2}
with the help of the method of differential equations \cite{DE}.
The corresponding results are expressed through so-called
two-dimensional harmonic polylogarithms which
generalize harmonic polylogarithms \cite{2dHPL}.

A reduction procedure that provides the possibility
to express any given Feynman integral to the master integrals was
also developed  in \cite{GR2}. It is based on
the Laporta's  observation that, when increasing the total dimension of
the denominator and numerator in Feynman integrals associated with the given graph,
the total number of IBP and Lorentz-invariance equations grows faster
than the number of independent Feynman integrals.
These techniques were successfully applied \cite{appl3j} to the Feynman
integrals with one leg off-shell contributing
to the process $e^+e^-\to 3$jets.

For another three-scale calculational problem, where
all four legs are on-shell and there is
a non-zero internal mass, a first analytical result was obtained
in \cite{S3} for the scalar master double box. These and other future similar
results will be used for calculations connected with Bhabha scattering
(see also \cite{Wert} for some steps in this direction).

It is believed that sooner or later we shall achieve the limit in
the process of analytical evaluation of Feynman integrals so that
we shall be forced to proceed only numerically.
(See, e.g., \cite{Pas} where this point of view has been emphasized.)
However the dramatic recent progress in the field of
analytical evaluation of Feynman integrals shows that we
have not yet exhausted our abilities in this direction.
Indeed, several powerful methods were developed last years.
To calculate the master integrals one can apply the technique of
MB integration and the method of differential equations mentioned above.
To construct appropriate recursive algorithms one can use recently
developed methods based on shifting dimension \cite{Tar} and
differential equations \cite{GR2} as well as a method based on non-recursive
solutions of recurrence relations \cite{Bai}.

One can also hope that new analytical results can be obtained
for many other classes of Feynman integrals depending on two and three
scales. In particular, the analytical evaluation of
any two-loop two-scale Feynman integral with two, three and four legs
looks quite possible.
In fact, when going to a higher level of calculational complexity,
one increases the number of loops, legs and independent variables.
As the experience with the double boxes with one leg off-shell has shown,
the crucial point is to introduce an appropriate class of functions.
In this example, these are two-dimensional harmonic polylogarithms
which turn out to be
adequate functions to express result for diagrams of the given family.
Presumably, when turning to a situation with one more kinematical invariant,
e.g. when one more leg is off-shell,
a natural procedure will be to introduce three-dimensional
harmonic polylogarithms, etc.

However, when the number of loops is increased one does not need to introduce
new functions, so that this transition means a `pure' calculational
complication.

\section{Triple boxes}

Let us now turn our attention to three-loop on-shell massless four-point diagrams
and consider dimensionally regularized massless
on-shell planar triple box diagram shown in Fig.~1.
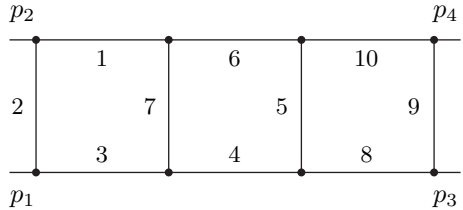
\begin {figure} [htbp]
\begin{picture}(170,60)(-25,-10)
\Line(-10,0)(0,0)
\Line(-10,50)(0,50)
\Line(160,0)(150,0)
\Line(150,0)(100,0)
\Line(150,50)(100,50)
\Line(160,50)(150,50)
\Line(0,0)(50,0)
\Line(50,0)(100,0)
\Line(100,50)(50,50)
\Line(50,50)(0,50)
\Line(0,50)(0,0)
\Line(50,0)(50,50)
\Line(100,0)(100,50)
\Line(150,0)(150,50)
\Vertex(0,0){1.5}
\Vertex(50,0){1.5}
\Vertex(100,0){1.5}
\Vertex(0,50){1.5}
\Vertex(50,50){1.5}
\Vertex(100,50){1.5}
\Vertex(150,0){1.5}
\Vertex(150,50){1.5}
\Text(-5,-10)[]{$p_1$}
\Text(155,-10)[]{$p_3$}
\Text(-5,60)[]{$p_2$}
\Text(155,60)[]{$p_4$}
\Text(25,43)[]{\small 1}
\Text(-7,25)[]{\small 2}
\Text(25,7)[]{\small 3}
\Text(75,7)[]{\small 4}
\Text(43,25)[]{\small 7}
\Text(93,25)[]{\small 5}
\Text(75,43)[]{\small 6}
\Text(125,7)[]{\small 8}
\Text(125,43)[]{\small 10}
\Text(143,25)[]{\small 9}
\end{picture}
\caption{Planar triple box diagram.}
\end{figure}
The general planar triple box Feynman integral without numerator
takes the form
\bea
T(a_1,\ldots,a_{10};s,t;\ep) &=&
\int\int\int \frac{\dd^dk \, \dd^dl \, \dd^dr}{(k^2)^{a_1}}
\nn \\ && \hspace*{-30mm}
\times \frac{1}{[(k+p_2)^2]^{a_2}[(k+p_1+p_2)^2]^{a_3}}
\nn \\ && \hspace*{-30mm} \times
\frac{1}{[(l+p_1+p_2)^2]^{a_4}[(r-l)^2]^{a_5}
(l^2)^{a_6}}
\nn \\ && \hspace*{-30mm} \times
\frac{1}{
 [(k-l)^2]^{a_7}[(r+p_1+p_2)^2]^{a_8} }
\nn \\ && \hspace*{-30mm}
\times \frac{1}{[(r+p_1+p_2+p_3)^2]^{a_9} (r^2)^{a_{10}} }
\, ,
\label{3box}
\eea
where $s=(p_1+p_2)^2$ and $t=(p_2+p_3)^2$ are Mandelstam variables, and
$k,l$ and $r$ are loop momenta.
Usual prescriptions $k^2=k^2+i 0, \; s=s+i 0$, etc. are implied.

By a straightforward generalization of two-loop manipulations,
one arrives \cite{S5} at a  sevenfold MB representation of (\ref{3box}).
In the case of the master triple box, $a_i=1,\;i=1,2,\ldots,10$, we have
\bea
T^{(0)}(s,t;\ep)\equiv T(1,\ldots,1;s,t;\ep)&&
\nn \\ &&  \hspace*{-37mm}
=\frac{\left(i\pi^{d/2} \right)^3}{\Gm(-2\ep)(-s)^{4+3\ep}}
\frac{1}{(2\pi i)^7}
\nn \\ &&  \hspace*{-52mm}\times
\int_{-i\infty}^{+i\infty}
\dd w \prod_{j=2}^7 \dd z_j
\left(\frac{t}{s} \right)^{w}
\frac{ \Gm(1 + w)\Gm(-w) }{\Gm(1 - 2 \ep + w - z_4) }
\nn \\ &&  \hspace*{-52mm}\times
\frac{ \Gm(-\ep + z_2)
  \Gm(-\ep + z_3) \Gm(1 + w - z_4)\Gm(-z_5) }
{ \Gm(1 + z_2 + z_4)
  \Gm(1 + z_3 + z_4)}
\nn \\ &&  \hspace*{-52mm}\times
\frac{\Gm(1 + \ep + z_4) \Gm(z_2 + z_4) \Gm(z_3 + z_4) \Gm(-z_6) }
{ \Gm(1 - z_5) \Gm(1 - z_6)  }
\nn \\ &&  \hspace*{-52mm}\times
\frac{\Gm(w + z_2 + z_3 + z_4 - z_7)\Gm(-z_2 - z_3 - z_4)
 }{\Gm(1 - 2 \ep + z_5 + z_6 + z_7)}
\nn \\ &&  \hspace*{-52mm}\times
 \Gm(1 + \ep + w - z_4 - z_5 - z_6 - z_7)
\Gm(1 + z_7)
\nn \\ &&  \hspace*{-52mm}\times
\Gm(-1 - \ep - z_5 - z_7) \Gm(-1 - \ep - z_6 - z_7)
\nn \\ &&  \hspace*{-52mm}\times
 \Gm(1 + z_5 + z_6 + z_7)  \Gm(-\ep - w - z_2 + z_5 + z_7)
\nn \\ &&  \hspace*{-52mm}\times
\Gm(2 + \ep + z_5 + z_6 + z_7)
\nn \\ &&  \hspace*{-52mm}\times
\Gm(-\ep - w - z_3 + z_6 + z_7)
\, .
\label{7MB0}
\eea

As a first step of three-loop calculations, let us consider
the evaluation \cite{S5} of
the leading power asymptotic behaviour
of (\ref{7MB0}) in the Regge limit $t/s \to 0$
This calculation demonstrates that a three-loop BFKL \cite{BFKL} analysis
is possible.

One can use the strategy of expansion by regions \cite{BS,SR,Sb}
which shows that
in the leading power only (1c-1c-1c) and (2c-2c-2c) regions contribute,
with the leading power $1/t$.
(See \cite{SR} and Chapter~8 of \cite{Sb} for definitions of these
contributions.)
The leading power (2c-2c-2c) contribution for the master planar triple box
takes the form
\bea
T^{(0),(2c-2c-2c)}(s,t;\ep)=
\int\int\int \frac{\dd^dk \, \dd^dl \, \dd^dr}{k^2 (k+p_2)^2}
&& \nn \\ && \hspace*{-67mm}
\times \frac{1}{(2 p_1 k+s)
(2 p_1 l+s) (r-l)^2
l^2 (k-l)^2}
\nn \\ && \hspace*{-67mm}
\times \frac{1}{ (2p_1 r+s) (r+p_2+\tilde{p})^2 r^2  }
\;,
\label{3box-c}
\eea
where $\tilde{p}$ is such that
$\tilde{p}^2=t, \; 2p_1 \tilde{p}=0, \; 2p_2 \tilde{p}=-t$.
The leading power (1c-1c-1c) contribution is obtained due to the
symmetry $\{1\lra 3,\;4\lra 6,\;8\lra 10\}$.

On the other hand, one can organize the calculational
procedure in such a way that the calculation of the Regge asymptotics is
a part of the calculation of the unexpanded triple box.
An analysis of the integrand shows that the key gamma functions that are
responsible  for the leading Regge behaviour are $\Gm(-\ep + z_{2,3})$ and
$\Gm(-1 - \ep - z_{6,5} - z_7)$.
The standard procedure of shifting contours and taking residues
can be applied. It results again in a sum of MB integrals where a
Laurent expansion of the integrand in $\ep$  is possible.
The final result for the Regge asymptotics of the planar
triple box is \cite{S5}:
\bea
T^{(0)}(s,t;\ep)=
-\frac{\left(i\pi^{d/2}
\e^{-\gm_{\rm E}\ep} \right)^3}{s^3 (-t)^{1+3\ep}}
&& \nn \\ &&  \hspace*{-50mm}
\times
\left\{
\frac{16}{9\ep^6}-\frac{5 L}{3 \ep^5}-\frac{3 \pi^2}{2 \ep^4}
\right.
- \left[\frac{11\pi^2}{12} L  + \frac{131 \zeta(3)}{9}
\right]\frac{1}{\ep^3}
\nn \\ &&  \hspace*{-43mm}
+  \left[\frac{49 \zeta(3)}{3} L- \frac{1411\pi^4}{1080}\right]
\frac{1}{\ep^2}
\nn \\ &&  \hspace*{-43mm}
-\left[\frac{503\pi^4 }{1440 }L  - \frac{73 \pi^2\zeta(3)}{4}
+ \frac{301 \zeta(5)}{15}\right]\frac{1}{\ep}
\nn \\ &&  \hspace*{-43mm}
+\left[\frac{223 \pi^2 \zeta(3)}{12}  + 149 \zeta(5)\right] L
\nn \\ &&  \hspace*{-43mm}
\left.
-\frac{624607 \pi^6 }{544320}  + \frac{167 \zeta(3)^2 }{9}
+ O(\ep) \right\} \;,
\label{Result}
\eea
where $L=\ln s/t$, $\gm_{\rm E}$ is the Euler constant
and $\zeta(z)$ is the Riemann zeta function.

For the analytic evaluation of (\ref{7MB0}), without expansion,
a similar procedure can be applied. Eventually, we arrive at
\be
T^{(0)}(s,t;\ep)=
-\frac{\left(i \pi^{d/2} \e^{-\gm_{\rm E}\ep} \right)^3 }{s^3 (-t)^{1+3\ep}}
\;
 \sum_{i=0}^6 \frac{ c_j(x,L)}{\ep^j}\;,
\ee
where, up to $1/\ep^2$ terms, we have
\bea
c_6=\frac{16}{9}\, , \;\;\;  c_5=-\frac{5 L}{3}\, , \;\;\;
c_4=-\frac{3 \pi^2}{2} \, , &&
\nn \\ &&  \hspace*{-63mm}
c_3=
3 (\Li{3}{-x} + L\, \Li{2}{-x})
\nn \\ &&  \hspace*{-57mm}
- \frac{3}{2}  (L^2+\pi^2) \ln(1 + x)
  -  \frac{11}{12}  L \pi^2 - \frac{131}{9}  \zeta(3) \,,
\nn \\ &&  \hspace*{-63mm}
c_2=
3(  S_{2,2}(-x)  + L \,  S_{1,2}(-x)) - 51 \Li{4}{-x}
\nn \\ &&  \hspace*{-57mm}
            + 3 \ln(1 + x) (\Li{3}{-x} + L\, \Li{2}{-x})
\nn \\ &&  \hspace*{-57mm}
- 37 L\,  \Li{3}{-x} -  \frac{23}{2}  L^2  \Li{2}{-x}
\nn \\ &&  \hspace*{-57mm}
- 8 \pi^2 \Li{2}{-x} - \frac{3}{4} (L^2 + \pi^2) \ln^2(1 + x)
\nn \\ &&  \hspace*{-57mm}
           - 3 \ln(1 + x) \zeta(3) + \frac{3}{2}  L^3 \ln(1 + x)
\nn \\ &&  \hspace*{-57mm}
+ L \pi^2\ln(1 + x)
            - \frac{1411}{1080}  \pi^4 + \frac{49}{3} L \zeta(3)\,.
\eea
Here $x=t/s$, so that $L=-\ln x$, and
\[
  S_{a,b}(z) = \frac{(-1)^{a+b-1}}{(a-1)! b!}
    \int_0^1 \frac{\ln^{a-1}(t)\ln^b(1-zt)}{t} \dd t
\]
are the generalized polylogarithms  \cite{GenPolyLog}.

The coefficient functions $c_i\,,\; i=2,\ldots,6,$  have
been confirmed \cite{BH1} by a numerical check
with the help of  numerical integration in
the space of alpha parameters  \cite{BH}. (This algorithm is based
on a procedure of resolution of singularities in the alpha
representation that was used in early papers on renormalization theory
(see, e.g., \cite{Hepp})
and in proofs of similar results for asymptotic expansions of Feynman
integrals in limits of momenta and masses (see, e.g., \cite{S6} and
Appendix~B of \cite{Sb}).

The present status of the calculation is as follows.
After integrating over some of the variables $z_i$,
$c_1$ and  $c_0$ are expressed through two-dimensional
MB integrals. In some of them, the last integration over a $z$-variable
can be  performed and resulting integrals over $w$
can be evaluated by closing the integration contour to the right
and taking residues at  $w=0,1,2,\ldots$. The corresponding
results are expressed in terms of harmonic polylogarithms  \cite{HPL} with
parameters $0$ and $1$.
In the rest of the contributions to $c_1$ and  $c_0$,
subintegrations cannot be done due to Barnes lemmas.
Hopefully, such two-dimensional MB integrals
can be analytically evaluated by means of `experimental mathematics'
similar to used in \cite{S3}, with results presumably  expressed
in terms of harmonic polylogarithms.

The procedure described above can be applied, in a similar way, to the
calculation of any massless planar on-shell triple box.

\vspace{0.2 cm}

{\em Acknowledgments.}
I am grateful to V.S.~Fadin and A.A.~Penin for helpful discussions
of perspectives of the three-loop BFKL analysis.
Thanks again to G.~Heinrich for numerical checks.

\end{document}